\documentclass[a4paper]{ESASPStyle}
\usepackage{epsfig}

\def\fig{.}
\def\dd{{\rm d}}

\begin{document}

\title{SOLA inversions for the core structure of solar-type stars}

\author{Sarbani Basu\inst{1} \and J{\o}rgen Christensen-Dalsgaard\inst{2} 
\and Michael J. Thompson\inst{3}} 
\institute{Astronomy Department, Yale University, 
New Haven CT, USA \and
Teoretisk Astrofysik Center, Danmarks Grundforskningsfond, and
Aarhus Universitet, {\AA}rhus, Denmark \and
Space \& Atmospheric Physics, Blackett Laboratory, 
Imperial College, London, U.K}

\maketitle 

\begin{abstract}

The Subtractive Optimally Localized Averages (SOLA) method,
developed and extensively used in helioseismology, is applied to
artificial data to obtain
measures of the 
sound speed inside a solar-type star. In contrast
to inversion methods which fit models to some aspect of the data, methods
such as SOLA provide an honest assessment of what can truly be
resolved using seismic data, without introducing additional assumptions such as
that the space of admissible stellar models is small. The resulting measures
obtained from SOLA inversion can subsequently be used to eliminate putative
stellar models. Here we present results of experiments to test the
reliability of SOLA inferences using solar models and models of
solar-type stars.

\end{abstract}

\section{Introduction}
  
Seismology of solar-type stars is expected in the not-too-distant future
to provide information of relevance for understanding stellar structure.
We need to develop tools that will allow us to use the full potential of the
information provided by the seismic observations of other stars.

Early attempts at inverting artificial sets of low-degree modes have given
mixed results. Gough \& Kosovichev (1993) and Roxburgh et al. (1998) inverted
the frequencies calculated for a 1.1 M$_\odot$ and a 0.8 M$_\odot$
star, respectively
to obtained very encouraging results.
More recently, Berthomieu et al. (2001) carried out a careful analysis of
the results to be expected in inversions of solar-like oscillations,
taking into account the expected errors in the COROT observations and
mode amplitudes in a 1.45~M$_\odot$ star; they concluded that reasonably
well-localised inversion is possible in the core of such a star.
On the other hand, Basu et al. (2001) failed
in their attempts at inverting a set of low-degree modes of a solar-type star.

One of the main differences between these efforts was the mode set used
for the inversions.
The Roxburgh et al. work used a very optimistic set:
it is unlikely that all those modes can be observed.
The Gough \& Kosovichev
mode set was somewhat more conservative.
Basu et al. (2001) used two conservative mode sets. The modes sets
and errors used in the different works are summarized in Table.~1.
In addition, unlike Gough \& Kosovichev and Roxburgh et al., Basu et al.
tried to invert for the sound speed $c$ using density $\rho$ as the second
variable, instead of inverting for $u\equiv p/\rho$, $p$ being the
pressure, with the helium abundance $Y$ as the second variable. It is
known that because density kernels (at fixed $c^2$)
have much larger amplitudes than $Y$ kernels (at fixed $u$),
a $c$ inversion is more difficult than a $u$ inversion.
The Basu et al. (2001) inversions were done under the assumption that
the mass, or radius, or both of these properties of the test star
was not known very well and hence this introduced additional\break
uncertainty.
Also we believe that there were numerical errors in the kernels used.

In this work we attempt to understand the differences in the results, in
particular we study the effect of the mode set and errors on the
results obtained. The mode sets used are the same as those used by 
Basu et al. (2001); the mode set of Gough \& Kosovichev (henceforth the GK set)
is used for comparison.
In order to be able to compare with older results
we try to invert for $u$ rather than $c^2$ which is the variable of
choice in solar inversions. To avoid additional uncertainties
due to uncertainty in the radius or the mass of the star under consideration,
in this work we confine our attention to solar models. Finally, we note that the
assumed errors are much larger than is typical in solar frequencies.

\begin{table}[bht]
  \caption{Mode sets considered and assumed standard deviation of
  errors. 
}
  \label{tab:tableone}
  \begin{center}
    \leavevmode
    \footnotesize
    \begin{tabular}[h]{lll}
      \hline \\[-5pt]
Paper & Mode set & Assumed \\[-1 pt]
      &          & errors\\[+5 pt]
      \hline \\[-5pt]
      Gough \& Kosovichev &  $l=0,1$, $n$=10-30 & $0.3\,\mu$Hz\\
(GK set) \hfill           & $l=2$, $n$=9-29  \\
\noalign{\smallskip}
Basu et al.\hfill    & $l=0$, $n=$11-27& $0.1\,\mu$Hz\\
Set 1 \hfill               & $l=1$, $n=$12-28  \\ 
                & $l=2$, $n=$15-27  \\
Basu et al.\hfill    & $l=0$, $n=$14-32& $0.3\,\mu$Hz$^\ast$\\
Set 2\hfill         & $l=1$, $n=$13-29  \\
                & $l=2$, $n=$15-30  \\
                & $l=3$, $n=$16-28 \\
\noalign{\smallskip}
Roxburgh et al.\hfill &     $l=0$, $n$=1-31 & $0.3\,\mu$Hz\\
                &     $l=1,2$, $n=$1-30 \\
                &     $l=3$, $n=$1-29 \\
      \hline \\
      \end{tabular}
  \end{center}
$^\ast${\it For Set~2, various errors are considered (see Fig.~5).}
\end{table}

\section{Models and inversion techniques}
\label{sec:mod}

Inversions for stellar structure are based on
linearizing the equations of stellar oscillations
around a known reference model. The relative differences in $u$ (i.e.
$\delta u/u$) and the difference in $Y$ ($\delta Y$)
between a star and a reference stellar model can be related to the
differences in the frequencies of the star and the model
($\delta\omega_i/\omega_i$),
\begin{equation}
{\delta\omega_i\over\omega_i}=\int K^i_{u,Y}(r)\dd r
+ \int K^i_{Y,u}(r)\dd r
+ {F_{\rm surf}(\omega_i)\over Q_i},
  \label{eq:freqdif}
\end{equation}
where $r$ is the normalized distance to the centre. The
index $i$ numbers the multiplets $(n,l)$.
The kernels $K_{c^2,\rho}^i$
and  $K_{\rho,c^2}^i$ are known functions of the reference model.
The term in $F_{\rm surf}(\omega_i)$
is the contribution from the uncertainties in the near-surface region
(e.g. Christensen-Dalsgaard \& Berthomieu 1991);
here $Q_i$ is the mode inertia, normalized by the inertia of a
radial mode of the same frequency.
In general, the right-hand side of equation (1) may also contain a term
${1\over 2}\delta\ln(M/R^3)$ to absorb scaling of the frequencies
with stellar mass $M$ and radius $R$ according to $(G M / R^3)^{1/2}$.

For linear inversion methods, the solution at a given point $r_0$ is
determined by a set of inversion coefficients $c_i(r_0)$, such
that the inferred value of, say, $\delta u/u$ is
\begin{equation}
\left < {\delta u\over u} (r_0)\right >=
\sum_i c_i(r_0){\delta\omega_i\over\omega_i}.
\label{eq:invcof}
\end{equation}
{}From the corresponding linear combination of
equations (\ref{eq:freqdif}) it follows that the
solution is characterized by {\it the averaging kernel},
obtained as
\begin{equation}
{\cal K} (r_0,r) = \sum_i c_i(r_0) K_{u,Y}^i(r) ,
\label{eq:avker}
\end{equation}
and also by the cross-term kernel:
\begin{equation}
{\cal C} (r_0,r) = \sum_i c_i(r_0) K_{Y,u}^i(r) \; ,
\label{eq:crosst}
\end{equation}
which measures the influence of the contribution from
$\delta Y$ on the inferred $\delta u/u$.

The surface term in equation~(\ref{eq:freqdif}) may be suppressed by
assuming that $F_{\rm surf}$ can be expanded in terms of polynomials
$\psi_\lambda$, and constraining the inversion coefficients to satisfy
$\sum_i c_i(r_0) Q_i^{-1} \psi_{\lambda}(\omega_i) = 0$,
$\lambda=0,1,...,\Lambda$.
(D\"appen et al. 1991).

The goal of the inversions is to obtain a localized averaging kernel,
while suppressing the contributions
from the cross term and the surface term
in the linear combination in equation~(\ref{eq:invcof}),
and limiting the error in the solution.
Also, ${\cal K}(r_0, r)$ must have unit integral
with respect to $r$.
If this can be achieved, then
\begin{equation}
\left < {\delta u\over u} (r_0)\right >\simeq
\int {\cal K} (r_0,r) {\delta u\over u} \dd r
\label{eq:avcb}
\end{equation}
defines a proper average of $\delta u/u$.

Here we use the Subtractive Optimally Localized Averages (SOLA)
method (Pijpers \& Thompson 1992, 1994) to determine the inversion
coefficients such that the
 averaging kernel is an approximation to
a given target ${\cal T}(r_0,r)$.
Details on the application of the SOLA technique to structure inversion
were given by, e.g.,
Rabello-Soares, Basu \& Christensen-Dalsgaard (1999).

The reference model used in this work is the solar Model S of
Christensen-Dalsgaard et al. (1996). This is a standard solar model.

The test model (the proxy star) we use is model MIX of
Basu, Pinsonneault \& Bahcall (2000). This is a non-standard model,
with an artificially mixed core. This model was selected because the
differences in $u$ with respect to the reference model
are large, somewhat along the
lines of what we expect for stars other than the Sun due to uncertainties
in their ages.

\section{Results}
\label{sec:res}

\subsection{Comparing $c^2$ and $u$ inversions}
\label{subsec:cu}

\begin{figure}[ht]
  \begin{center}
    \epsfig{file=\fig/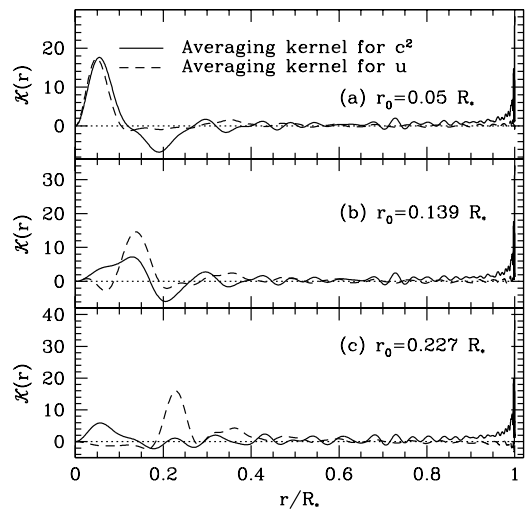, width=7cm}
  \end{center}
\caption{ \small Averaging kernels for sound-speed inversions (continuous) and $u$ inversions
(dashed) for a few target radii. The mode set used was Set~2 with a 
uniform error of $0.3\,\mu$Hz.
\label{fig:figavk}}
\end{figure}

\begin{figure}[ht]
  \begin{center}
    \epsfig{file=\fig/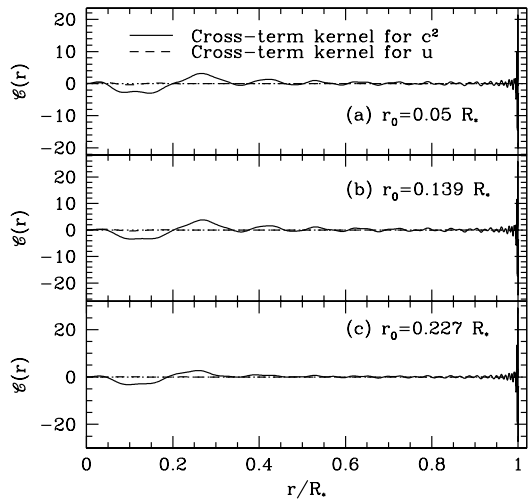, width=7cm}
  \end{center}
\caption{\it Cross-term kernels for the inversions.
\label{fig:figcr}}
\end{figure}

Fig.~\ref{fig:figavk} shows a comparison of
averaging kernels obtained for $c^2$ and $u$ inversions. The data set
used was Set~2, with uniform errors of $0.3\,\mu$Hz. Note that the
$c^2$ averaging kernels have a lot of surface structure, which
has a detrimental effect on the inversion results. The mode set needs
to be expanded in order to get a cleaner averaging kernel.
Fig.~\ref{fig:figcr} shows a comparison of
the cross-term kernels obtained for $c^2$ and $u$ inversions.
One can see that with the mode set used (Set~2), we cannot successfully suppress the
cross term in a $c^2$ inversion. In contrast, the cross term for the
$u$ inversions is very small.

\subsection{Effect of mode set}
\label{subsec:set}

\begin{figure}[ht]
  \begin{center}
    \epsfig{file=\fig/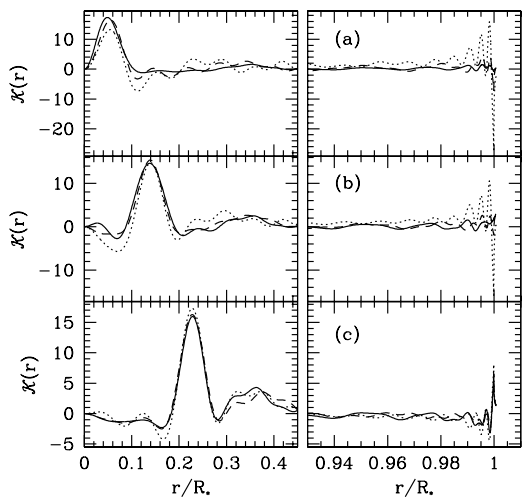, width=7cm}
  \end{center}
\caption{ \small The averaging kernels obtained with Set~1 (dotted lines), Set~2 (continuous
lines) and GK (dashed lines) mode sets at 0.05 R$_\star$ (Panel a), 
0.139 R$_\star$ (Panel b) and 0.227 R$_\star$ (Panel c). The near surface structure
of these kernels is shown in the panels to the right.
\label{fig:avkcomp}}
\end{figure}

\begin{figure}[ht]
  \begin{center}
    \epsfig{file=\fig/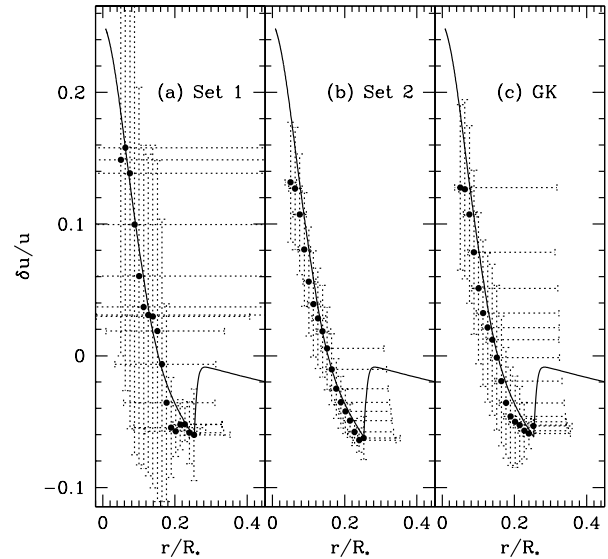, width=8.00cm}
  \end{center}
\caption{\small  Results of $u$ inversion for
Set~1 (panel a), Set~2 (panel b) and GK (panel c). The  continuous
line is the exact difference between the models.
Inversion results are plotted at the corresponding target radius.
The vertical error bars  show the $1$-$\sigma$ uncertainty in the inversion
result, obtained by propagating the mode uncertainties through the inversion
process. The horizontal error bars
provide an indication of the resolution of the inversion;
they extend from the first to the third
quartile point of the averaging kernels.
A uniform error of 0.3$\,\mu$Hz was assumed for each set.
\label{fig:all}}
\end{figure}

Fig.~\ref{fig:avkcomp} shows
a comparison of the
averaging kernels  obtained for $u$ inversions
using mode sets Set~1, Set~2 and GK. A uniform error of 0.3$\,\mu$Hz was
assumed for all sets. Note that the Set~1 averaging kernels have
structure at the surface. Set~2 and GK kernels are much cleaner.
Set~1 also results in very large cross-term kernels. It should be
noted that MOLA inversions can give somewhat better averaging kernels.

Fig.~\ref{fig:all} shows the inversion results
for the three sets. The horizontal error bars are an indication of
the resolution of the inversion. One can see that the Set~1 results have
very large errors, as well as poor resolution. Set~2 and GK give very good
results for a substantial portion of the core.

\subsection{Effect of mode errors}
\label{subsec:errors}

\begin{figure}[ht]
  \begin{center}
    \epsfig{file=\fig/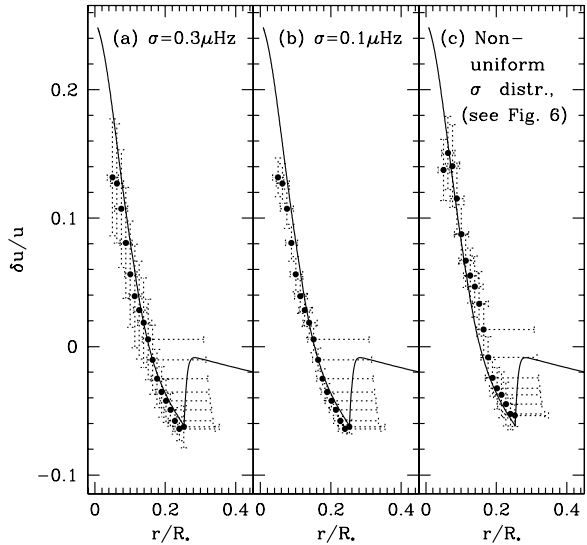, width=7.8cm}
  \end{center}
\caption{ \small Results of $u$ inversion with Set~2. Different error distributions
were used for the inversions whose results are shown in the different
panels.
\label{fig:inv}}
\end{figure}

\begin{figure}[ht]
  \begin{center}
    \epsfig{file=\fig/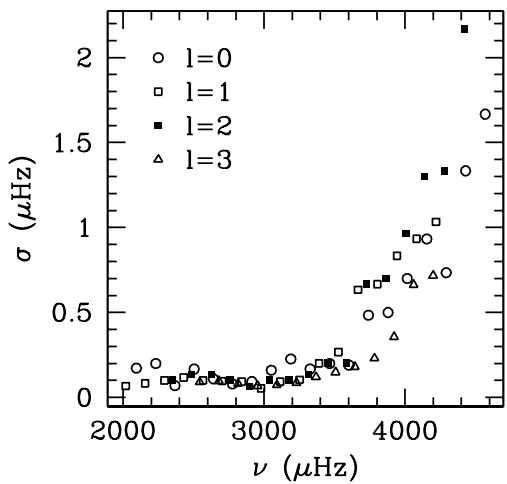, width=7cm}
  \end{center}
\caption{\small  The distribution of errors used
for the inversions shown in Fig.~5(c). The errors
were scaled from the error distribution of solar
oscillation frequencies so that the low-frequency part of the
distribution is around $0.1\,\mu$Hz.
\label{fig:error}}
\end{figure}

A reduction in mode errors helps the inversions.
The effects can be felt in two ways: one could improve the resolution, while keeping the
error on the solution the same, or, one could keep the averaging kernel the same
and decrease the error on the solution. Fig.~\ref{fig:inv} shows the inversion
results for Set~2 for three different error-distributions: (1) uniform errors
of 0.3$\,\mu $Hz, (2) uniform errors of $0.1\,\mu $Hz, and (3) an error-distribution
which mimics the distribution of errors in solar oscillations frequencies. The
distribution was scaled to be around $0.1\,\mu$Hz
at low frequency (see Fig.~\ref{fig:error}
for the error distribution).
The reduction of errors gives a more dramatic
improvement for Set~1 results. If the errors are reduced by a factor of 5,
the surface structure of the averaging kernels can be removed
and the cross-term kernels can be made much smaller.

\section{Discussion}
\label{sec:disc}

In common with other investigations of the resolving power of the low-degree
modes expected from observations of solar-like oscillations in distant
stars, we find that with realistic assumptions about the expected mode
set and observational errors it is possible to achieve some limited
resolution in the core of the model.
With our present proxy star this, for example, shows clear evidence
for the core mixing as reflected in the steep rise in $\delta u/u$
towards the centre.

The results depend strongly on the choice of variables in the inversion.
For helioseismic inversion, a common choice is to express the inverse
problem in terms of adiabatic sound speed and density.
With the limited set of modes available in the stellar case,
the quality of the inversion for the sound speed suffers from
the need to suppress the contribution from the density.
A preferable choice of variables
is the pair $(u, Y)$;
here the contribution from $Y$ is small and essentially confined to
the outer layers, where helium is partly ionized.

Not surprisingly, the results show substantial dependence on the
choice of mode set, although the precise manner in which this
happens is not obvious and needs further study.
It is important here also to include more realistic estimates
of the expected properties of the modes and observations, as
has been attempted by Berthomieu et al. (2001).

In analyses of evolved stars it is possible that modes with partial
g-mode character may be observed; some 
evidence for this
may already have been found in the subgiant $\eta$ Bootis
(e.g. Christensen-Dalsgaard, Bedding \& Kjeldsen 1995).
Such modes would have a
highly beneficial impact on inversions for the stellar core, because the
averaging kernels could exploit the localized trapping of the eigenfunctions.

\begin{acknowledgements}
This work was supported by the Danish National Research Foundation
through the establishment of the Theoretical Astrophysics Center,
and by the UK Particle Physics and Astronomy Research Council.
\end{acknowledgements}

\end{document}